\documentclass{jpsj-suppl}

\title{Mean-Field Calculation Based on Proton-Neutron Mixed Energy Density Functionals}

\author{
Koichi~\textsc{Sato}$^1$, Jacek~\textsc{Dobaczewski}$^{2,3,4}$,
Takashi~\textsc{Nakatsukasa}$^{1,5}$, and Wojciech~\textsc{Satu{\l}a}$^{2,4}$
}

\inst{
$^1$RIKEN Nishina Center,2-1 Hirosawa, Wako, Saitama 351-0198, Japan\\
$^2$Institute of Theoretical Physics, Faculty of Physics,University of Warsaw, ul. Pasteura 5, 02-093 Warsaw, Poland\\
$^3$Department of Physics, PO Box 35 (YFL), FI-40014
University of Jyv{\"a}skyl{\"a}, Finland \\
$^4$Helsinki Institute of Physics, P.O. Box 64, FI-00014 University of Helsinki, Finland \\
$^5$Center for Computational Sciences, University of Tsukuba,
Tennodai 1-1-1, Tsukuba 305-8577, Japan
}

\email{satok@ribf.riken.jp}

\recdate{,2014}

\abst{
We have performed calculations based on the Skyrme energy density functional (EDF) that includes arbitrary mixing
between protons and neutrons. In this framework, single-particle states are generalized as mixtures of
proton and neutron components. The model assumes that the Skyrme EDF is invariant under the rotation in
isospin space and the Coulomb force is the only source of the isospin symmetry breaking. To control the isospin of the system,
we employ the isocranking method, which is analogous to the standard cranking approach used for describing high-spin states. Here, we
present results of the isocranking calculations performed for the isobaric analog states in $A=40$ and $A=54$ nuclei.
}

\kword{proton-neutron mixing, energy density functionals, isobaric
analog states}

\begin{document}
\maketitle

\section{Introduction}

 The proton-neutron (p-n) pairing is a long-standing open problem in nuclear physics,
and its possible relations to various nuclear phenomena have been widely discussed \cite{[Per04]}.
However, in spite of the recent impressive experimental progress and
theoretical studies,
the understanding of the p-n pairing is still unsatisfactory.
To address this problem, we use the nuclear density functional approach.
Our ultimate goal is to develop a superfluid symmetry-unrestricted
energy-density-functional (EDF) approach including the p-n mixing
both in the pairing and particle-hole (p-h) channels. Indeed, in accordance with
fundamental self-consistency requirements of the Hartree--Fock(--Bogoliubov) (HF(B)) equations,
any generalization of quasiparticle states as mixtures of proton and neutron
components must be necessarily accompanied by, somewhat less intuitive, mixing of proton and neutron single-particle
(s.p.) wave functions.

Recently, as the first step in developing the superfluid EDF theory including the p-n mixing, by extending
the codes {\sc HFODD}~\cite{[Sch12]} and {\sc HFBTHO}~\cite{[Sto13]}, in Refs.~\cite{[Sat13c],[She14]} we have developed a
s.p.\  EDF formalism including the p-n mixing in the p-h channel.
In this p-n mixing calculation, we applied the so-called isocranking method by adding the isocranking
term to the Hamiltonian:
$\hat{h'}=\hat {h}-\vec{\lambda} \cdot \hat{\vec{t}}$,
where $\hat {\vec{t}}$ is the isospin operator.

Our model is based on a local Skyrme EDF extended to include the p-n
mixing by following the general rule given by Perli\'nska {\it et al.}~\cite{[Per04]}.
Starting from the local density matrix $\rho({\bf r},tt')$ ($t$ and $t^\prime$ are the
isospin indices), we built  the isoscalar $\rho_0({\bf r})= \sum_{tt'}
\rho({\bf r},tt')\hat{\tau}^0_{t't}$ and isovector
$\vec{\rho}({\bf r})= \sum_{tt'} \rho({\bf r},tt')\hat{\vec{\tau}}_{t't}$ densities by contracting $\rho({\bf r},tt')$
with the isospin identity matrix $\hat{\tau}_0$ and isospin Pauli matrices $\hat{\vec{\tau}}$,
respectively. The isoscalar density $\rho_0({\bf r})$ and isovector $z$ component are the
sum and difference of neutron and proton densities, respectively. These densities are included in the conventional
EDF calculations. The $x$ and $y$ components of the isovector densities are new elements, which we take into account
to extend the EDFs, and which are nonzero only for the p-n mixed s.p.\ states.

In the following, we present selected numerical results obtained in $A=40$ and $A=54$ nuclei
for the SkM* EDF parameters set\cite{[Bar82]}.
The applications are divided into two classes, without and with the Coulomb interaction.
The reason is that we have extended our Skyrme EDFs such that they are invariant under the
rotation in the isospin space. If the Coulomb interaction is
switched off, the total and s.p.\ energies should be independent of the isospin direction of the system,
which allows us to validate numerical implementation of the code.
The Coulomb interaction, when included, is calculated exactly both in the direct and exchange channels.

\section{Numerical Results and Discussion}

It is well known that the isospin symmetry is only weakly broken in atomic nuclei and the concepts of the isospin
conservation and isospin quantum number prevail even in the presence of the Coulomb interaction. In order to
control the approximate isospin conservation we have employed, as already mentioned, the isocranking method, which corresponds
to the lowest-order isospin projection.
We parametrize the isocranking frequency $\vec{\lambda}$ as
\begin{equation}
\vec \lambda=(\lambda\sin \theta,0,\lambda\cos\theta)
=(\lambda^{\prime}\sin \theta^\prime,0,\lambda^{\prime}\cos\theta^\prime+\lambda_{\rm off}). \label{eq:shiftedsemicircle}
\end{equation}
Even in the calculation with the Coulomb term, our Hamiltonian is
invariant under rotation about the $T_z$ axis. Therefore, we set $\lambda_y=0$
and consider the isocranking only in the $T_x - T_z$ plane.
The procedure of the isocranking calculations
is as follows \cite{[Sat13c]}. First, we perform the standard Hartree-Fock (HF) calculation for the
isoaligned ($|T_z|=T$) states and thus we find the corresponding neutron and proton Fermi energies
$\lambda_n$ and $\lambda_p$.
Next, we determine values of $\lambda_{\rm off}$ and $\lambda^{\prime}$ as
\begin{equation}
 (\lambda_{\rm off}, \lambda^\prime)=
\frac{1}{2}(\lambda_{np}^{T_z=T} +\lambda_{np}^{T_z=-T},
\lambda_{np}^{T_z=T} -\lambda_{np}^{T_z=-T} ),
\label{eq:lambda}
\end{equation}
where $\lambda_{np}^{T_z=\pm T}\equiv \lambda_n-\lambda_p$ is the difference
of the neutron and
proton Fermi energies in the $T_z=\pm T$ isobars.
Finally, we vary the tilting angle $\theta^\prime$ with $\lambda^{\prime}$ and $\lambda_{\rm off}$ fixed.
In the calculations without the Coulomb interaction,
we set $\lambda_{\rm off}=0$ and $\theta=\theta^\prime$.
Due to the Coulomb interaction, the s.p.\ levels vary in function of  $\langle
\hat T_z \rangle$.
Therefore, in varying the tilting angle $\theta$ from 0$^\circ$ to 180$^\circ$,
level crossings may take place.
The choice of $\vec \lambda$ in Eq.~(\ref{eq:shiftedsemicircle}) helps avoiding the level
crossings and smooths a way to obtain
the isobaric analogue states (IASs) from  $\theta^\prime=0^\circ$ to 180$^\circ$.

\begin{figure}[tbp]
\begin{center}
\includegraphics[height=5.3cm]{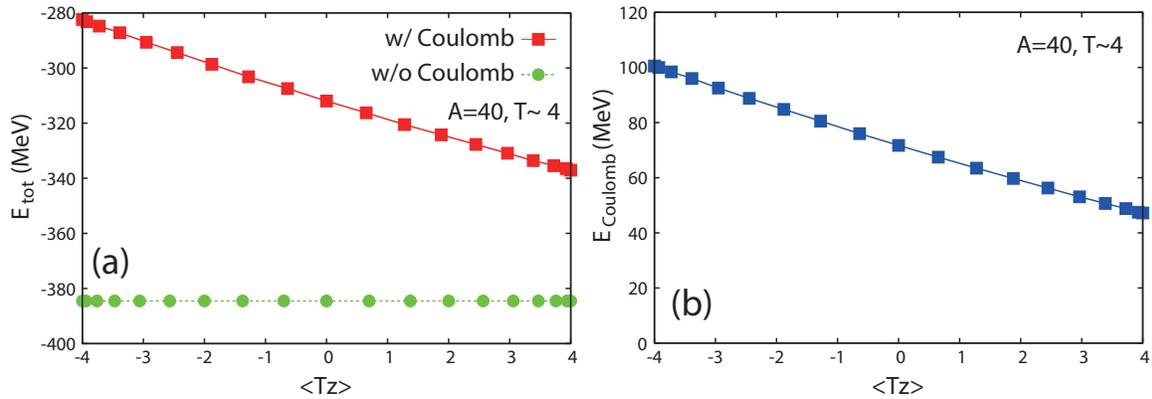}
\caption{(a) Total energies of $T \approx 4$ IASs in
 $A=40$ isobars calculated with and  without the Coulomb interaction in function
 of $\langle \hat T_z\rangle$.
(b) Coulomb energies of $T \approx 4$ IASs in $A=40$ in function of $\langle \hat T_z\rangle$.
The results of isocranking calculations for every $10^\circ$ of $\theta^\prime$ between
$\theta^\prime =0^\circ$ and $180^\circ$ are plotted.
}
\label{fig:s_040_energy}
\end{center}
\end{figure}

In Fig.~\ref{fig:s_040_energy}(a), we show the total energies of the $T\simeq 4$ IASs in $A=40$ isobars calculated with and without the
Coulomb interaction.
We have used  $(\lambda_{\rm off}, \lambda^\prime)=(0, 12.5)$ MeV and
$(\lambda_{\rm off}, \lambda^\prime)=(-6.8, 13.6)$ MeV
for the calculations without and with the Coulomb interaction, respectively.
They are determined from
the difference of the proton and neutron Fermi energies
$\lambda_{n}-\lambda_{p}$
in the standard HF solution for $^{40}$S and $^{40}$Cr.
When the Coulomb interaction is switched off, our EDF is invariant under the isospin rotation.
The total energy calculated without the Coulomb interaction
is independent of the direction of the
isospin, which constitutes a test of the code.

When the Coulomb interaction is switched on, the total energy depends
on  the expectation value $\langle \hat T_z \rangle$. One can see that the total energy depends
on $\langle \hat T_z \rangle$ almost linearly. The effect comes predominantly
from the Coulomb energy which exhibits almost the same dependence on
$\langle \hat T_z \rangle$ as that of the total energy as shown in Fig.~\ref{fig:s_040_energy}(a).
Its linearity results from proportionality of the Coulomb energy to $Z^2 = T_z^2-AT_z+A^2/4$, that
clearly enhances the linear term by a factor of $A$ as compared to the quadratic term
for small $T_z$.

\begin{figure}[tbp]
\begin{center}
\includegraphics[height=5.3cm]{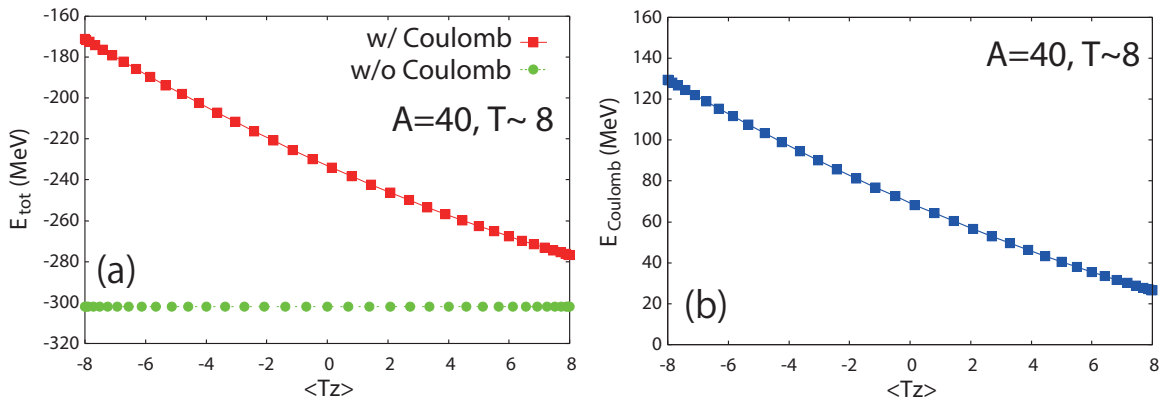}
\caption{Same as Fig.~\ref{fig:s_040_energy} but calculated for the $T\simeq 8$ IASs in $A=40$ isobars.
The results of isocranking calculations for every $5^\circ$ of $\theta^\prime$ between
$\theta^\prime =0^\circ$ and $180^\circ$ are plotted.
}
\label{fig:mg040_energy}
\end{center}
\end{figure}

In Fig.~\ref{fig:mg040_energy},
we show the same results as in Fig.~\ref{fig:s_040_energy},
but calculated for the $T\simeq 8$ states in the $A=40$ isobars.
For these calculations, we have used
$(\lambda_{\rm off}, \lambda^\prime)=(0, 27.4)$ MeV and
$(\lambda_{\rm off}, \lambda^\prime)=(-6.0, 28.6)$ MeV
for the calculations without and with the Coulomb interaction, respectively.
The values are determined from the standard HF ground state solutions in $^{40}$Mg and $^{40}$Ni.
Again, the total energy calculated without the Coulomb interaction is
independent on $\langle \hat T_z \rangle$.
However, in Fig.~\ref{fig:mg040_energy}(b) one can see traces of the quadratic dependence of the Coulomb energy
on $\langle \hat T_z \rangle$, although the contribution from the linear term is dominant.

\begin{figure}[tb]
\begin{center}
\includegraphics[width=0.75\textwidth]{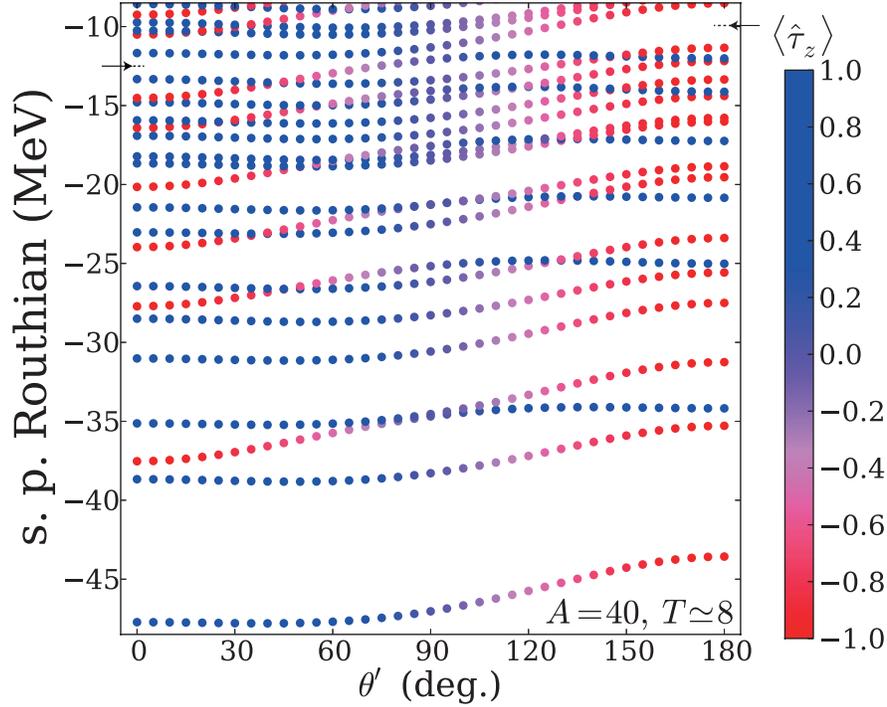}
\caption{
Single-particle Routhians of the $T\simeq 8$ states in $A=40$
 isobars calculated with the Coulomb interaction included. The arrows at
 the upper left and upper right indicate the positions of the Fermi energies at
 $\theta^\prime =0^\circ$ and 180$^\circ$, respectively.}
\label{fig:mg040routhian}
\end{center}
\end{figure}

In Fig.~\ref{fig:mg040routhian}, we plot the expectation values of the s.p.\ Routhian
$\hat h^\prime=\hat h - \vec \lambda \cdot \hat \vec t$,
calculated for the $T\simeq 8$ IASs with $A=40$.
At $\theta^\prime=0$, the Fermi surface appears around $-12.5$\,MeV, below which
14 neutron and 6 proton orbitals are occupied.
The s.p.\ Routhians vary as  functions of $\theta^\prime$, and, unlike in the case of $A=48$ \cite{[Sat13c]}, there is no
large shell gap above the Fermi surface.
Nevertheless, with our choice of $\vec \lambda$, the level crossings are avoided.
While the s.p.\ states are pure proton or neutron states at $\theta=0^{\circ}$ and 180$^{\circ}$,
which means that the $|T_z|=T$ states are nothing but the standard HF states without the p-n mixing,
at all other tilting angles, the s.p. states are p-n mixed.
In particular, the proton and neutron components are almost equally
mixed at $\theta^\prime=90^\circ$, which corresponds to $T_z \approx 0$.

\begin{figure}[tbp]
\begin{minipage}{0.48\hsize}
\begin{center}
\includegraphics[height=5.2cm]{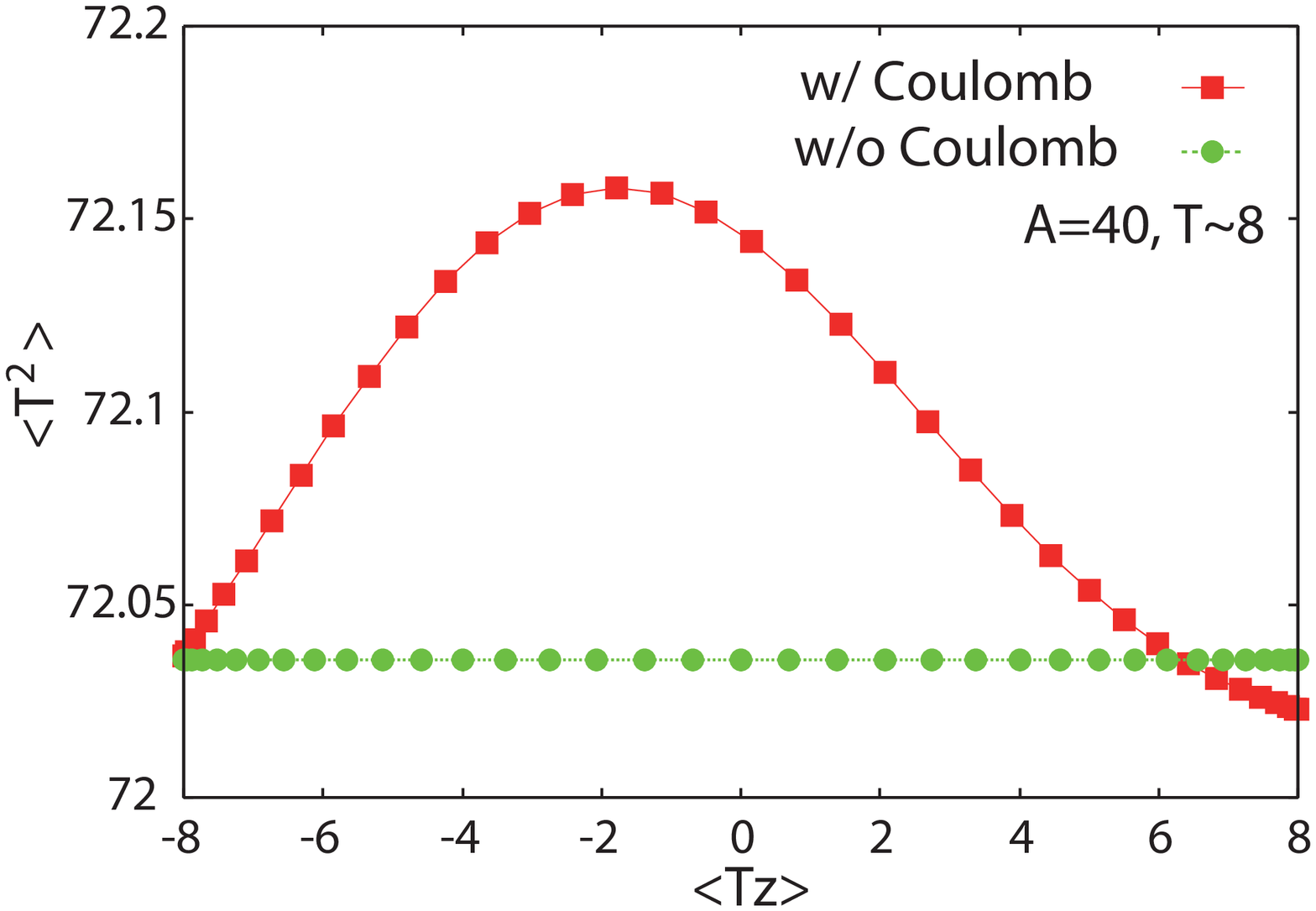}
\caption{$\langle \hat T^2 \rangle$ calculated for $T\simeq 8$ IASs
 with and without the Coulomb interaction.}
\label{fig:mg040_t2}
\end{center}
\end{minipage}\hspace*{0.5cm}
\begin{minipage}{0.48\hsize}
\begin{center}
\includegraphics[height=5.2cm]{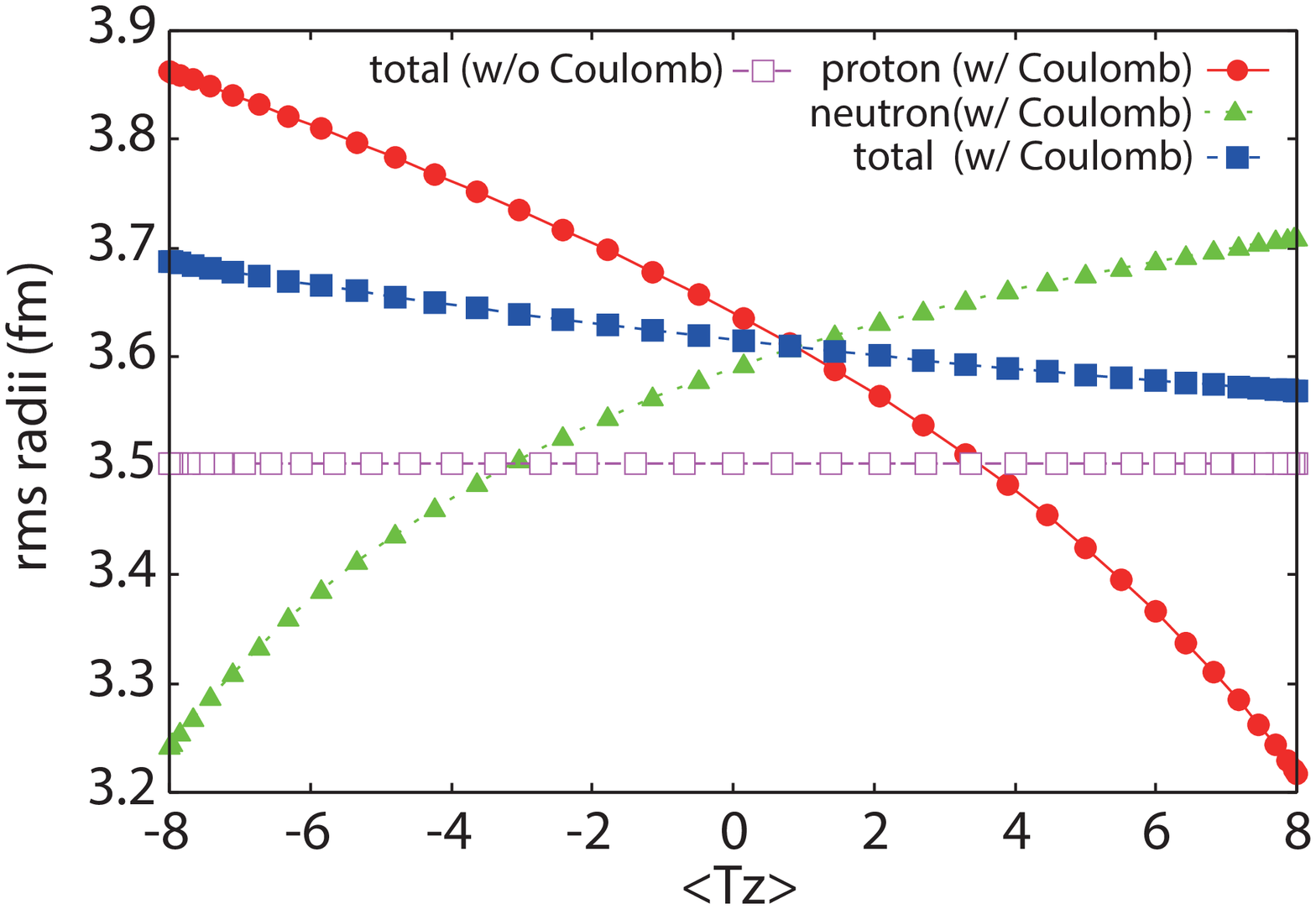}
\caption{Root-mean-square radii of the $T =8$ IASs with
 $A=54$ as functions of $\langle \hat T_z\rangle$. }
\label{fig:mg040_rms}
\end{center}
\end{minipage}
\end{figure}

Fig.~\ref{fig:mg040_t2} shows
the expectation values of $\langle \hat T^2 \rangle$
calculated for the $T \simeq 8$ states in $A=40$ isobars.
In case of rigorous isospin conservation one should obtain $\langle \hat T^2 \rangle$=72.
The Coulomb interaction breaks the isospin symmetry and gives a deviation from this value.
However, even in the case without the Coulomb interaction, the calculated
$\langle \hat T^2 \rangle$ deviates from the exact value 72 due to the spurious isospin mixing within
the mean-field approximation \cite{[Eng70],[Aue83],[Sat09a]}. Note that around $T_z=8$ the spurious deviation is
even larger than in the case with the Coulomb interaction.

\begin{figure}[tbp]
\begin{minipage}{0.48\hsize}
\begin{center}
\vspace*{-1.2cm}
\includegraphics[height=5.1cm]{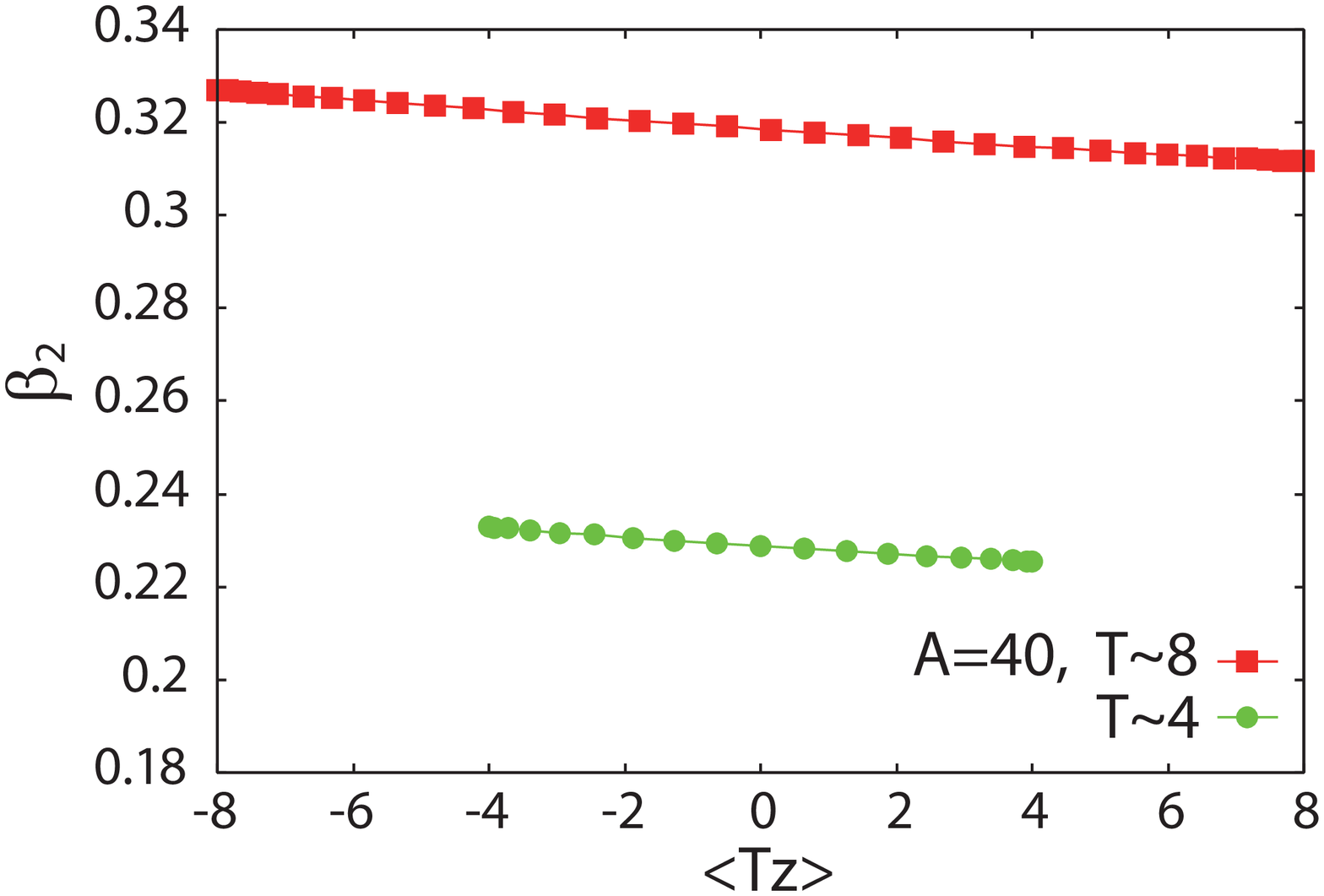}
\vspace*{0.3cm}
\caption{Quadrupole deformation $\beta_{2}$ calculated for the $T\simeq 4$
 and $T \simeq 8$ IASs in $A=40$ isobars with the Coulomb interaction included.}
\label{fig:a040_b20}
\end{center}
\end{minipage}\hspace*{0.5cm}
\begin{minipage}{0.48\hsize}
\begin{center}
\includegraphics[height=5.1cm]{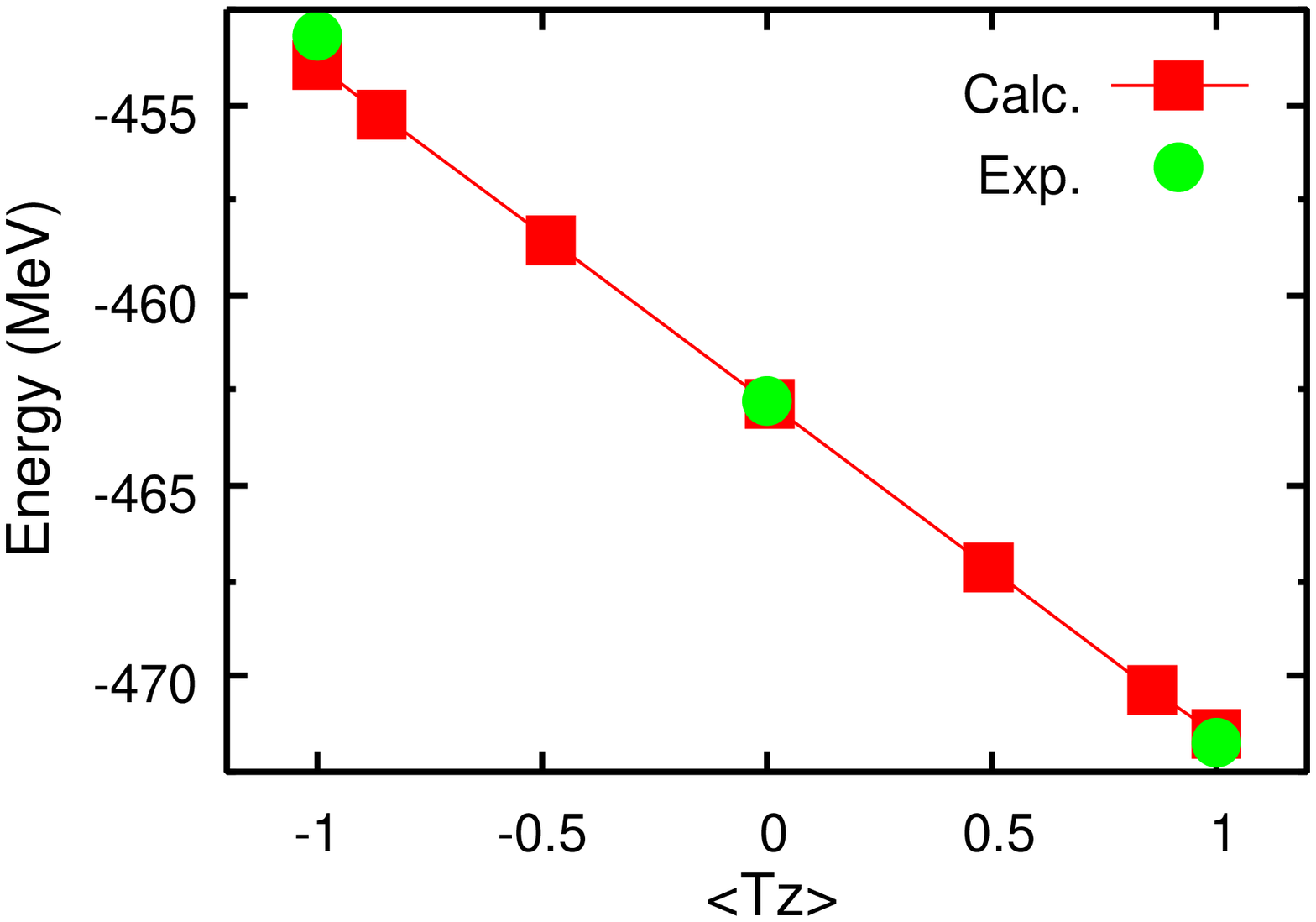}
\caption{Energies of $T \simeq 1$ isobaric analog states in $A=54$ isobars in comparison with
the experimental data \cite{[Brookhaven]}.
The results of isocranking calculations for every $30^\circ$ of $\theta^\prime$ between
$\theta^\prime =0^\circ$ and $180^\circ$ are plotted.}
\label{fig:fe054}
\end{center}
\end{minipage}
\end{figure}

Fig.~\ref{fig:mg040_rms} shows the proton, neutron and total root-mean-square (rms) radii
calculated with the Coulomb interaction for the $T \simeq 8$ states in $A=40$ isobars,
together with the total rms radius calculated without the Coulomb interaction.
The neutron (proton) rms radius increases with increasing (decreasing)
$\langle \hat T_z \rangle$, that is, increasing the neutron (proton) components.
With the Coulomb interaction, the total rms radius increases with increasing the proton components
due to the Coulomb repulsion among protons. Without the Coulomb
interaction, it stays constant as a function of $\langle \hat T_z \rangle$.
In Fig.~\ref{fig:a040_b20}, we depict the quadrupole deformation
parameter $\beta_2$ calculated for the $T\simeq
4$ and $T\simeq 8$ IASs in $A=40$ isobars. In both of the IAS
chains, the quadrupole deformation $\beta_2$ is nearly constant, which illustrates
the fact that the s.p.\ configuration for all IASs stays the same.

In the $A=4n$ nuclei, such as the $A$=40 systems discussed above,
even-$T$ states are the ground states of even-even nuclei and their IASs.
We also performed calculations for $A=4n+2$ nuclei, in which odd-$T$
states are the ground states of even-even nuclei.
As an example of those calculations,  in Fig.~\ref{fig:fe054},
we depict the calculated energies of the $T=1$
triplet in $A=54$ isobars in comparison with the experimental data.
Here, the $I=0^+, T \simeq |T_z|=1$ states are the ground states
of $^{54}$Fe and $^{54}$Ni and are described by the standard HF solutions
without the p-n mixing.
On the other hand, the $T_z=0$ IAS, the lowest $I=0^+$ state in $^{54}$Co,
is obtained by the isocranking calculation,
and it consists of the p-n mixed s.p. states.
It is gratifying to see that both the energy of the $T_z=0$ state
as well as those of the $|T_z|=1$ states are well reproduced by
the theory. It is worth stressing that the $T_z=0$ IAS in an odd-odd nucleus
is described here by means of a single time-even Slater determinant.
This is at variance with single-reference p-n unmixed EDF models, wherein such states
do not exist at all~\cite{[Sat10]}.

\section{Concluding Remarks}

In this work, we have solved the generalized self-consistent Skyrme EDF equations
including the arbitrary mixing between protons and neutrons in the p-h channel.
The values of the total isospin and its $T_x$ and $T_z$ components of the
system were controlled by the isocranking method, which is analogous to the tilted-axis cranking calculation for high-spin
states. We have performed isocranking calculations for
even-$T$ $A$=40 IASs and odd-$T$ $A$=54 IASs demonstrating
that the single-reference EDF approach including p-n mixing is
capable of quantitatively describing the IASs both in the even-even as well as in the odd-odd nuclei.

Here, we have used the isocranking method to control the isospin,
which is a simple linear constraint method. In the code {\sc HFODD},
we have also implemented a more sophisticated method for
optimizing the constraint \cite{[Sat13c]}, known as the augmented Lagrange method,
and we applied it to
calculate the excitation energies of the $T \simeq 0,2,4,6,$, and $8$ states in $^{48}$Cr.

Recently,
by extending an axially-symmetric Skyrme HFB code {\sc HFBTHO}~\cite{[Sto13]},
another Skyrme EDF code with the p-n mixing
has been developed in Ref.\ \cite{[She14]}.
We performed benchmark tests by comparing the results of the
isocranking calculations obtained with the codes {\sc HFBTHO} and {\sc HFODD},
and we obtained an excellent agreement.

As discussed in Ref. \cite{[Sat10]}, there is spurious isospin mixing inherent to the mean-field approach.
In order to remove this spurious mixing, one needs to perform the isospin projection and the subsequent
Coulomb rediagonalization. The implementation of the isospin projection
into our p-n EDF code is now in progress.

\section*{Acknowledgments}
This work is partly supported by JSPS KAKENHI (Grants No. 25287065), NCN (Contract No. 2012/07/B/ST2/03907),
by the THEXO JRA within the EU-FP7-IA project ENSAR (No.\ 262010), by the ERANET-NuPNET grant
SARFEN of the Polish National Centre for Research and Development,
and by the Academy of Finland and University of Jyv\"askyl\"a within the FIDIPRO programme.
The numerical calculations were carried out on a SR16000 computer at the
Yukawa Institute for Theoretical Physics in Kyoto University
and at the RIKEN Integrated Cluster of Clusters (RICC) facility.


\end{document}